\documentclass[twocolumn,superscriptaddress,aps,prl,showpacs,amsmath,amsfonts,amssymb,reprint]{revtex4-1}
\usepackage{graphicx}

\makeatletter
\usepackage[usenames]{color}
\usepackage{calc}
\usepackage[update,prepend]{epstopdf}
\makeatother

\begin{document}

\title{Observation of mode-locked spatial laser solitons}

\author{F. Gustave}

\affiliation{Institut Non Lin\'{e}aire de Nice, UMR 7335 CNRS, 06560 Valbonne, France}

\author{N. Radwell}

\affiliation{SUPA and Department of Physics and Astronomy, University of Glasgow,
Glasgow G12 8QQ, UK}

\author{C. McIntyre}

\affiliation{SUPA and Department of Physics, University of Strathclyde, Glasgow
G4 ONG, Scotland, UK}

\author{J. P. Toomey}

\affiliation{MQ Photonics Research Centre, Dept. of Physics \& Astronomy, Macquarie
University, Sydney, 2109, Australia}

\author{D. M. Kane}

\affiliation{MQ Photonics Research Centre, Dept. of Physics \& Astronomy, Macquarie
University, Sydney, 2109, Australia}

\author{S. Barland}

\affiliation{Institut Non Lin\'{e}aire de Nice, UMR 7335 CNRS, 06560 Valbonne, France}

\author{W. J. Firth}

\author{G.-L. Oppo}

\author{T. Ackemann \footnote{To whom correspondence should be addressed.}\email[]{thorsten.ackemann@strath.ac.uk}}

\affiliation{SUPA and Department of Physics, University of Strathclyde, Glasgow
G4 ONG, Scotland, UK}

\date{V5, \today}
\date{\today}

\begin{abstract}
A stable nonlinear wave packet, self-localized in all three dimensions,
is an intriguing and much sought after object in  nonlinear science in general and in nonlinear
photonics in particular. We report on the experimental
observation of mode-locked spatial laser solitons in a vertical-cavity
surface-emitting laser with frequency-selective feedback from an external
cavity. These spontaneously emerging and long-term stable spatio-temporal structures have a pulse length shorter than the cavity round trip time and may pave the way to completely independent cavity light bullets.
\end{abstract}
\maketitle

Wave packets in general, and light wave packets in particular, do not remain confined
to small regions in space or time, but have the natural tendency to
broaden. In space this is due to diffraction and in time this is due
to dispersion, at least in a medium. Although localization can be
achieved by confining potentials (e.g.\ optical fibres and photonic
crystals in optics), it was always the vision of researchers to obtain confinement by self-action. This is why the concept of solitary waves,
or, more loosely speaking, of solitons received a lot of attention
over the last decades. A soliton is a wavepacket in which the tendency
to broaden is balanced by nonlinearities. Spatial and temporal solitons
in one or two dimensions are known in many fields of science  \cite{remoissenet94,stegeman99,grelu12,ackemann09a,akhmediev05,akhmediev08,purwins10}.
    %\cite{hasegawa02,stegeman99,grelu12,ackemann09a,purwins10,vanag07,bel12,niemela90,umbanhowar96,akhmediev05,akhmediev08}.
    %, e.g.\ optics \cite{hasegawa02,stegeman99,grelu12,ackemann09a}, gas discharges
    %\cite{purwins10}, chemistry \cite{vanag07}, ecology \cite{bel12},
    %hydrodynamics \cite{niemela90}, and granular media \cite{umbanhowar96}.
There was also early interest in three dimensional (3D) self-localization,
not  least as a model for elementary particles \cite{derrick64}. However,
early results \cite{derrick64} were discouraging as they indicated that stationary 3D localized states are not stable in a broad
class of nonlinear wave equations. Significant interest continued
in the theoretical and mathematical literature (e.g.\ \cite{muratov96,muratov00,bode02,barashenkov02,graham07,amin10,thiele13}), but to our knowledge there is only evidence for 3D self-organized periodic patterns \cite{bansagi11}, but not for 3D self-localization.
%3D oscillons were shown to be unstable \cite{barashenkov02}.

In optics, spatio-temporal localization of light, i.e.\ a `bullet'
of light being confined in all three spatial dimensions (and time,
because it is propagating), was considered as early as 1990 in the
framework of a 3D Nonlinear Schr\"{o}dinger-equation (NLSE) \cite{silberberg90},
but realized to be unstable due to the well known blow-up experienced
for cubic nonlinearities in more than one dimension \cite{rasmussen86,berge98}.
There were several proposals of how to stabilize multi-dimensional solitons
    %by various means
\cite{kolokolov73,enns92,akhmediev93,berge98,veretenov00,wise02,mihalache04,malomed05,berge98,panagiotopoulos15,veretenov16}.
    %\cite{kolokolov73,enns92,akhmediev93,berge98,veretenov00,mihalache04,berge98,panagiotopoulos15,veretenov16}
    %(see \cite{wise02,malomed05} for a review).
Spatio-temporal compression \cite{koprinkov00,gaeta01,eisenberg01},
meta-stable 2D spatio-temporal solitons \cite{liu99a,wise02} and
meta-stable discrete light bullets \cite{minardi10,eilenberger13}
were observed in pioneering experiments, the latter constituting
the closest realization of a stable light bullet up until now. However,
these quasi-conservative bullets are only stable over a few characteristic
lengths and are long-term unstable due to losses and Raman-shifts.

This makes it attractive to look for solitons in dissipative optical
systems in which losses are compensated by driving \cite{akhmediev05}.
    % (see \cite{akhmediev05} for an overview).
Indeed solitons can exist in 2D with a cubic
nonlinearity within a cavity \cite{firth96a} or with parametric driving
\cite{barashenkov02}. Dissipative bullets were analyzed theoretically
in the cubic-quintic Ginzburg-Landau model \cite{grelu05,skarka06,mihalache06,akhmediev07,mihalache07},
intra-cavity second harmonic generation \cite{tlidi99a}, two-level
models for driven cavities \cite{brambilla04}, optical parametric
oscillators \cite{veretenov09}, solid-state lasers \cite{renninger14}
and semiconductor lasers \cite{javaloyes16}. In particular, the latter
seem to be very promising as advances in semiconductor laser technology
yielded 2D spatial solitons \cite{tanguy08,genevet08,elsass10a,ackemann09a}
as well as 1D temporal solitons and mode-locking \cite{paschotta02,genevet09,marconi14,gustave15}. The latter matches a huge literature on temporal dissipative solitons
in mode-locked solid-state and fiber lasers, see e.g.\ \cite{grelu12}.
Hence, it seems to be attractive to look at mode-locking spatial laser
solitons to achieve 3D localization.

\begin{figure*}[t]
\centering \includegraphics[width=\textwidth]{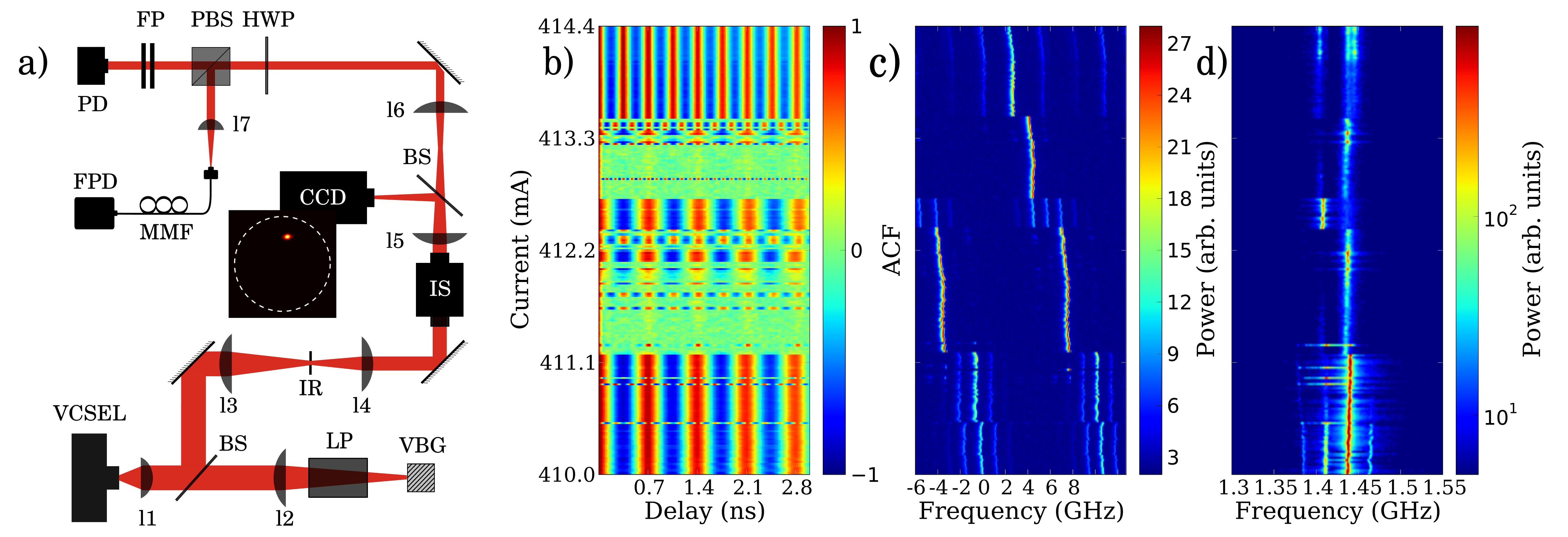}
 \protect\caption{(Color online) a) Scheme of experimental setup. The inset illustrates spatial localization by showing the soliton within the much larger pumped VCSEL aperture indicated by the dotted white line. L lens, BS beam
sampler, LP linear polarizer, VBG volume Bragg grating, IR iris, IS
optical isolator, CCD charge-coupled device camera, HWP half-wave
plate, PBS polarizing beam splitter, FP Fabry-Perot interferometer
(free spectral range (FSR) of 12 GHz and a Finesse of about 40),
PD photodector, FPD fast photodetector. b-d) Correlation functions
and spectra obtained for different currents (scanning down) in the
bistable range, b) Autocorrelation function, c) Optical spectra obtained
from FP (the overall blue-shift with decreasing current is due to
the reduced thermal load; approximately 1.7 FSR are displayed to show
unwrapped spectra in spite of the blue-shift) and d) RF-spectra of
the intensity signal on the FPD from an electrical spectrum analyzer
(ESA) around the fundamental round-trip frequency.}
\label{fig:spectra}
\end{figure*}

We are addressing the issue in a vertical-cavity surface-emitting
laser (VCSEL) with frequency-selective feedback (FSF) from an external cavity
\cite{tanguy08,radwell09}, and demonstrate self-pulsing of spatial
solitons based on the locking of three external cavity modes. This constitutes
already an important novel experimental observation in optical self-organization
and nonlinear photonics.
Furthermore, these spatio-temporal structures result from self-organization
on scales shorter than the spatial extent of the compound
cavity in all three dimensions of space and therefore pave the way to
    %fully independent
light bullets.

The experimental system (Fig.~\ref{fig:spectra}a) consists of a VCSEL with an InGaAs gain
material emitting around 980 nm \cite{radwell09} and an aperture
diameter of 200~$\mu$m. The laser is coupled to a self-imaging external
cavity closed by a volume Bragg grating (VBG) as a frequency filter. The VBG has a reflectivity peak with 99\% reflection at around
981~nm with a full width at half maximum (FWHM) of 30~GHz. The round-trip time in the
external cavity is about 0.7~ns. The output is monitored locally with a fast
photo-detector (bandwidth 9.5 GHz) followed by an AC-coupled RF-amplifier
(bandwidth 15 GHz) and is then digitized by a real time oscilloscope
at 100 GS/s (bandwidth 23 GHz).  A linear polarizer forces the polarization of the solitons to be vertical
ensuring 10\% outcoupling via the Fresnel reflection of an intra-cavity
beam sampler. The detected area is broader than the soliton, i.e.\ the detector is only sensitive to variations of total power and not to breathing or jittering modes. The optical spectrum is monitored with a plano-planar
scanning Fabry-Perot interferometer.
    %(FP).
Changing the current does not only change the gain, but the dominant effect is actually to change the detuning between VCSEL resonance and VBG resonance via Ohmic heating (e.g.\ \cite{jimenez16,ackemann16}). This change of detuning is our main control parameter. For small enough detuning, optical bistability occurs due to a carrier mediated, dispersive nonlinearity \cite{gustave16s,jimenez16,ackemann16}.  Situations
of single-mode operation in the external cavity (corresponding to
cw spatial solitons \cite{radwell09}) and multi-mode operation  can be realized. Some self-pulsing was found in preliminary investigations \cite{ackemann10a,ackemann16,toomey14}. The
measurements reported in this letter demonstrate long-term mode-locking of solitons with an excellent
phase coherence. We stress that the mode-locked states reported here are bistable
\cite{gustave16s}, i.e.\ it can be either present or absent for fixed values of the parameters, a typical feature of localized dissipative structures.
    %We also stress that the time averaged images on the CCD (inset in Fig.~\ref{fig:spectra}a)
    %show the same shape and size for the stationary and pulsating solitons,
    %thus confirming an amplitude modulation.

\begin{figure}
\centering \includegraphics[width=45mm]{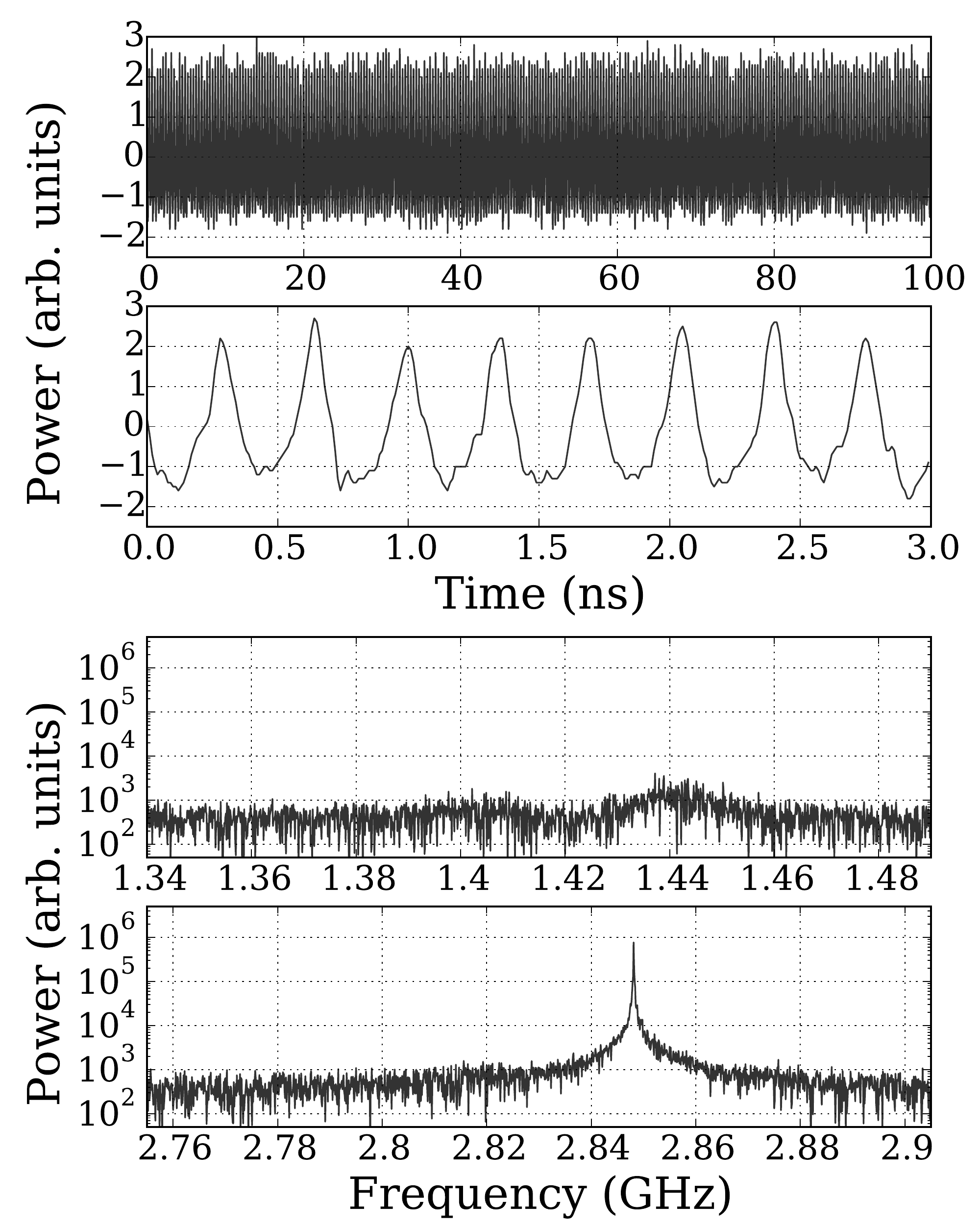} \includegraphics[width=40mm]{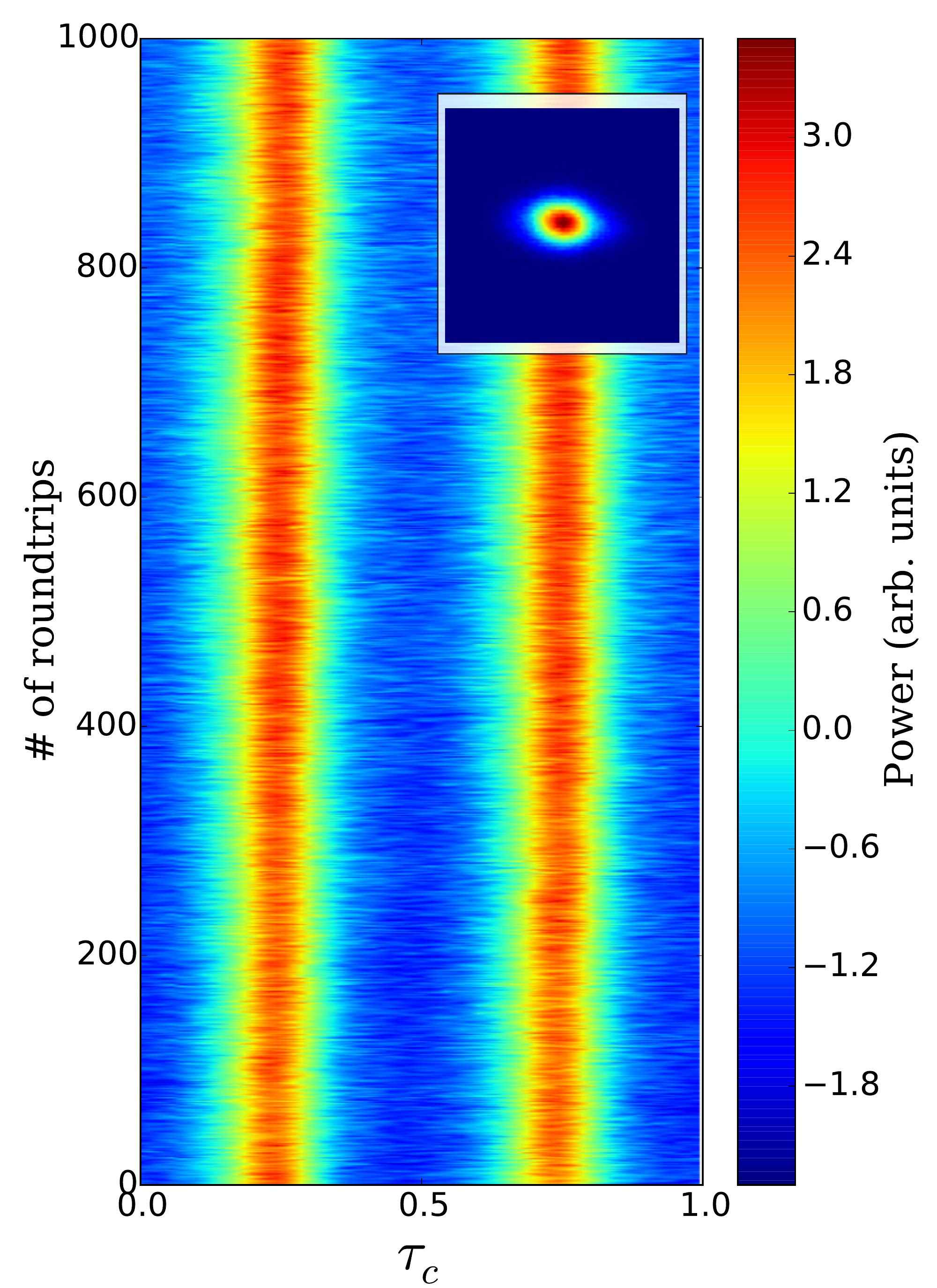}
\protect\caption{(Color online) Example of harmonic mode-locking, $I=413.7$~mA. Left panel, from
top to bottom: Envelope of time series,  zoom on pulses (the AC-coupled RF-amplifier removes the DC component
and hence the signal has positive and negative amplitudes), RF-spectra
obtained from a numerical fast Fourier transform (FFT) of the recorded
time series around the fundamental round-trip frequency, RF-spectrum
around harmonic. The right panel shows space-time representations
of the dynamics, the horizontal axis displays time within a round-trip
(space-like) and the vertical axis the number of round-trips. The inset shows the time averaged intensity profile of the pulsing soliton in a square with a size of 55~$\mu$m.}
\label{fig:HML}
\end{figure}

In the realization discussed here (VCSEL substrate temperature 70.5$^{\circ}$C), the soliton switches
on at 414.4 mA, as the current is ramped up slowly in steps of 0.02~mA. After a value of 418 mA is reached, the current  is
ramped down again and the soliton is sustained below the switch-on
threshold to a switch-off threshold below 409.9~mA. In the current range investigated in
Figs.~\ref{fig:spectra}b-d, the system is bistable,
i.e.\ the soliton coexists with the non-lasing state as expected for
a localized dissipative structure. Depending on current, the autocorrelation function (ACF,
Fig.~\ref{fig:spectra}b) shows distinct recurrences
at multiples and fractions of the cavity round-trip time of 0.7 ns
as well as very unstructured regimes. Comparison with the optical
spectra in Fig.~\ref{fig:spectra}c shows that regular
recurrence is typically associated with multi-mode dynamics
    % whereas the regimes with irregular ACF are associated with dominantly single-mode emission.
providing a first indication of self-pulsing via mode-locking.
This is supported by the radio-frequency (RF) spectra (Fig.~\ref{fig:spectra}d), showing a pronounced
peak at the cavity round-trip frequency around 1.4 GHz for the regions
of  regular ACF recurrence at 410-411.1 mA, 412.1 mA and around 412.5
mA. We are going to discuss typical examples for three
different manifestations of regular self-pulsing spatial solitons below.

\begin{figure}
\centering \includegraphics[width=45mm]{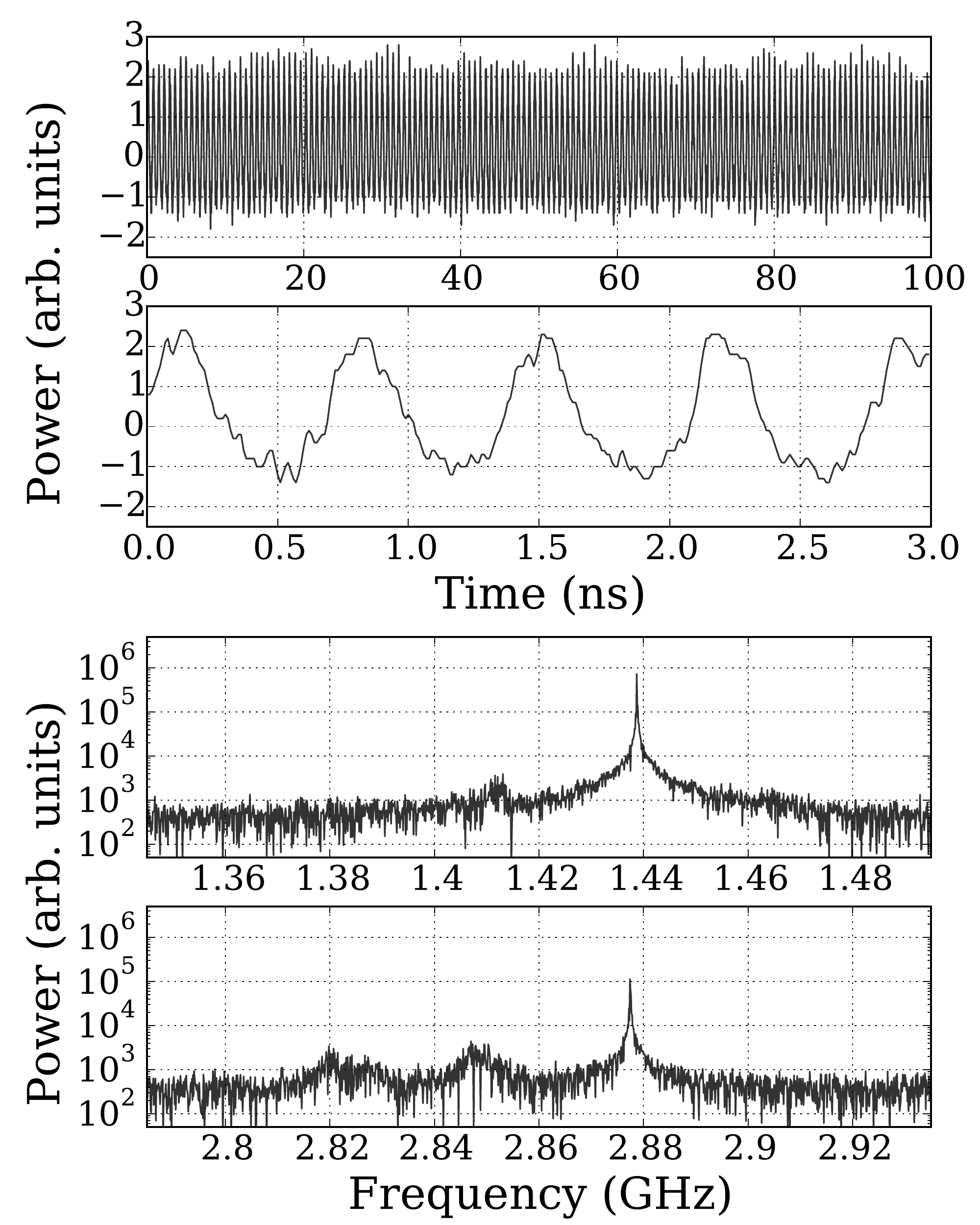} \includegraphics[width=40mm]{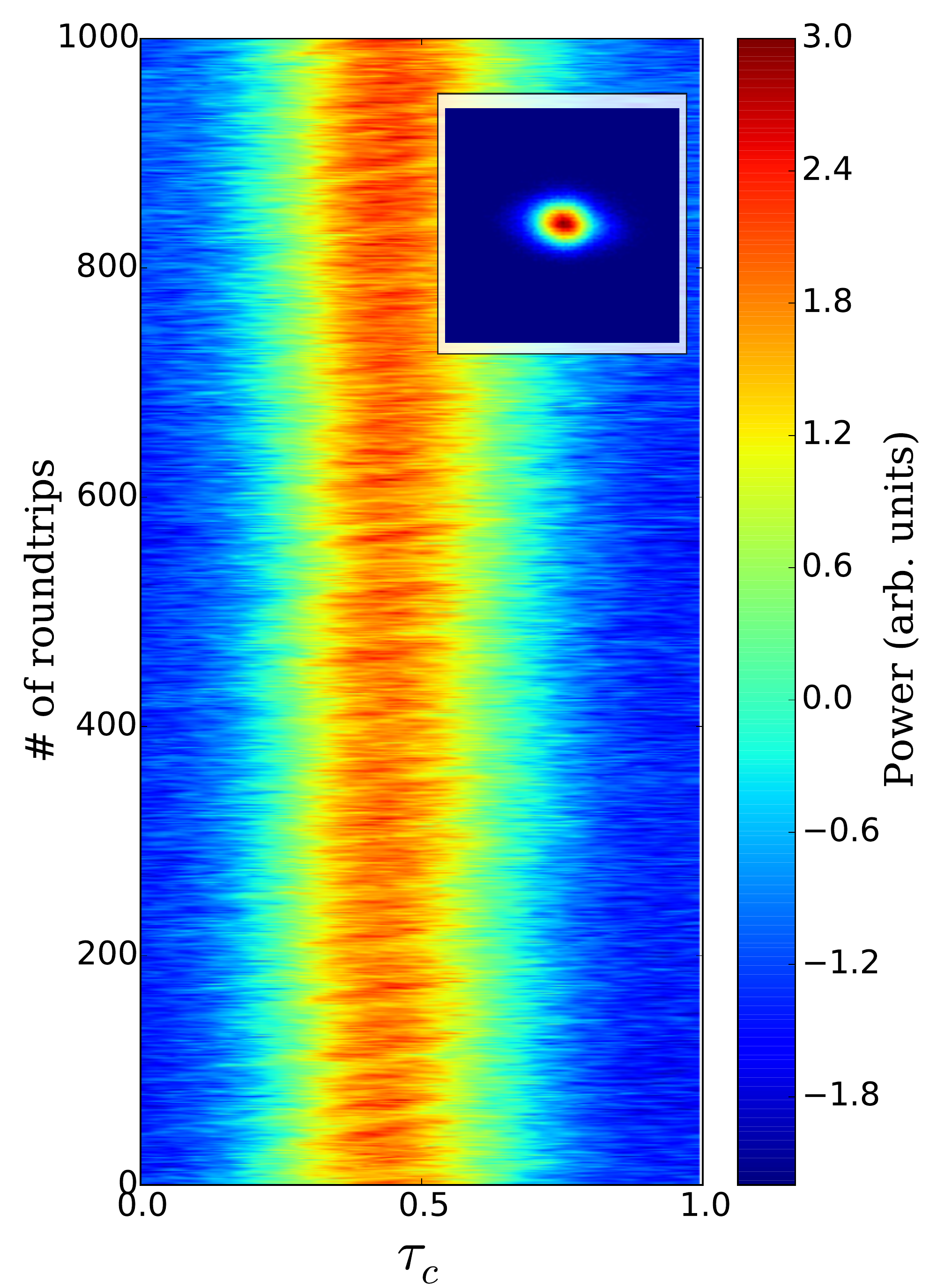}
\protect\caption{(Color online) Example of fundamental mode-locking, $I=410.6$~mA. Explanation
of subpanels as in Fig.~\ref{fig:HML}.}
\label{fig:FML}
\end{figure}

Fig.~\ref{fig:HML} depicts the situation at a current
of 413.7 mA. A fairly regular sequence of pulses is emitted with a
pulse separation of 351 ps (standard deviation (SDV) 10 ps). The dominant RF feature is found
just below 2.85~GHz. In the terminology of mode-locking, this corresponds to
\emph{harmonic mode-locking}, i.e.\ at two times the fundamental round-trip
frequency. Indeed there are two pulses within the cavity at the same
time as  demonstrated by the space-time representation in
the right panel of Fig.~\ref{fig:HML}. The horizontal
axis displays time within a round-trip frame (space-like, \cite{giocomelli96})
and the vertical axis displays slow time developing with the number
of round-trips. The optical spectrum in Fig.~\ref{fig:spectra}c
confirms that the excited modes are separated by two FSR of the external
cavity. There are three modes excited with two further faint side
modes. The average pulse width (FWHM) measured in the time domain is 112 ps (SDV 9 ps). This is in agreement with a numerical estimation of 116~ps obtained from the locking of three modes with the relative amplitudes as observed in the optical spectrum  \cite{gustave16s}.
    %The excellent quality of the mode-locking of these three modes is further demonstrated by the extreme narrowness
    %of the RF peak which is at the resolution limit given
    %by the observation time of 1/40 $\mu$s = 25 kHz.
The width of the RF peak is at the resolution limit given by the observation time of 1/40 $\mu$s = 25 kHz. This evidences excellent phase synchronization. The envelope is also fairly stable over long
time scales. The observations provide evidence of robust harmonic mode-locking of spatial solitons. We stress that the mode-locking prevails over much longer scales (seconds to minutes) than the recorded time span of 40 $\mu$s, which corresponds already to about 57000 round-trips. Its stability is limited by drifts in the setup changing the detuning conditions, as there is no active stabilization of cavity length.

Fig.~\ref{fig:FML} illustrates the situation at a
lower current of 410.6 mA. Again there is a fairly stable sequence
of pulses, but the pulse separation is now 695~ps (SDV 31 ps) and the
fundamental repetition frequency is at 1.438 GHz. The optical spectrum
in Fig.~\ref{fig:spectra}c shows again three main
modes with two much fainter sidemodes. They are separated by only
one FSR of the external cavity. The measured pulse duration is 270 (SDV 23 ps). This is again in good agreement with an estimation of 250~ps stemming from a numerical superposition of three modes
    %with the amplitudes as indicated by the experimental optical spectra
\cite{gustave16s}.
As the RF-amplifier in the experiment is AC-coupled, we do not have
direct access to the modulation depth, but the numerical superposition of three
modes indicates an effective valley-pulse peak modulation depth of 91-94\% and thus provides indirect confirmation of a high modulation depth also in experiment \cite{gustave16s}. There is only one pulse per round-trip.
This constitutes \emph{fundamental mode-locking}. The fundamental RF peak
is still resolution limited confirming excellent phase
correlation.  Again, the envelope
is fairly stable. There are weak indications of satellite peaks around
the fundamental RF peak but they are suppressed by nearly three orders
of magnitude.

\begin{figure}
\centering \includegraphics[width=45mm]{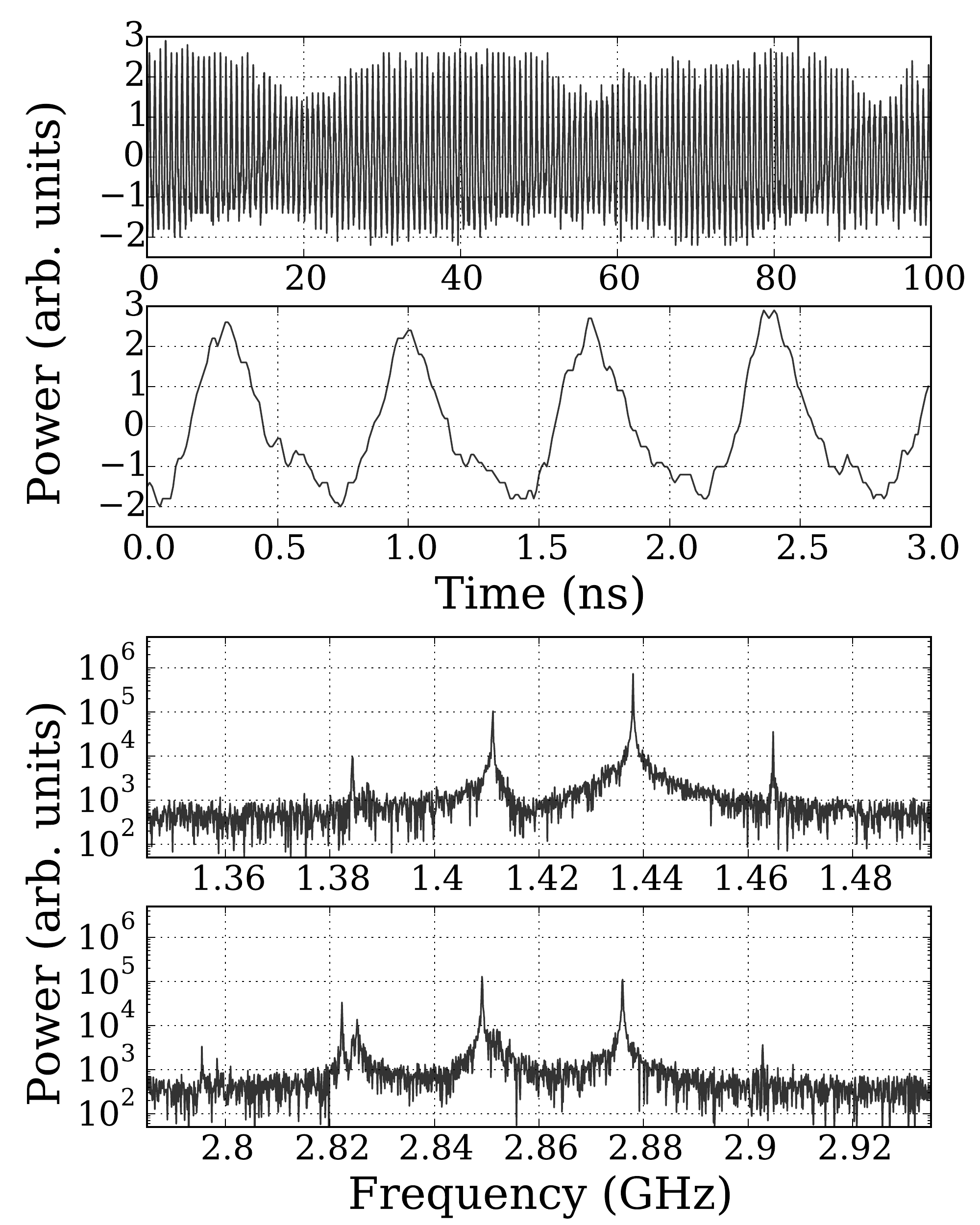} \includegraphics[width=40mm]{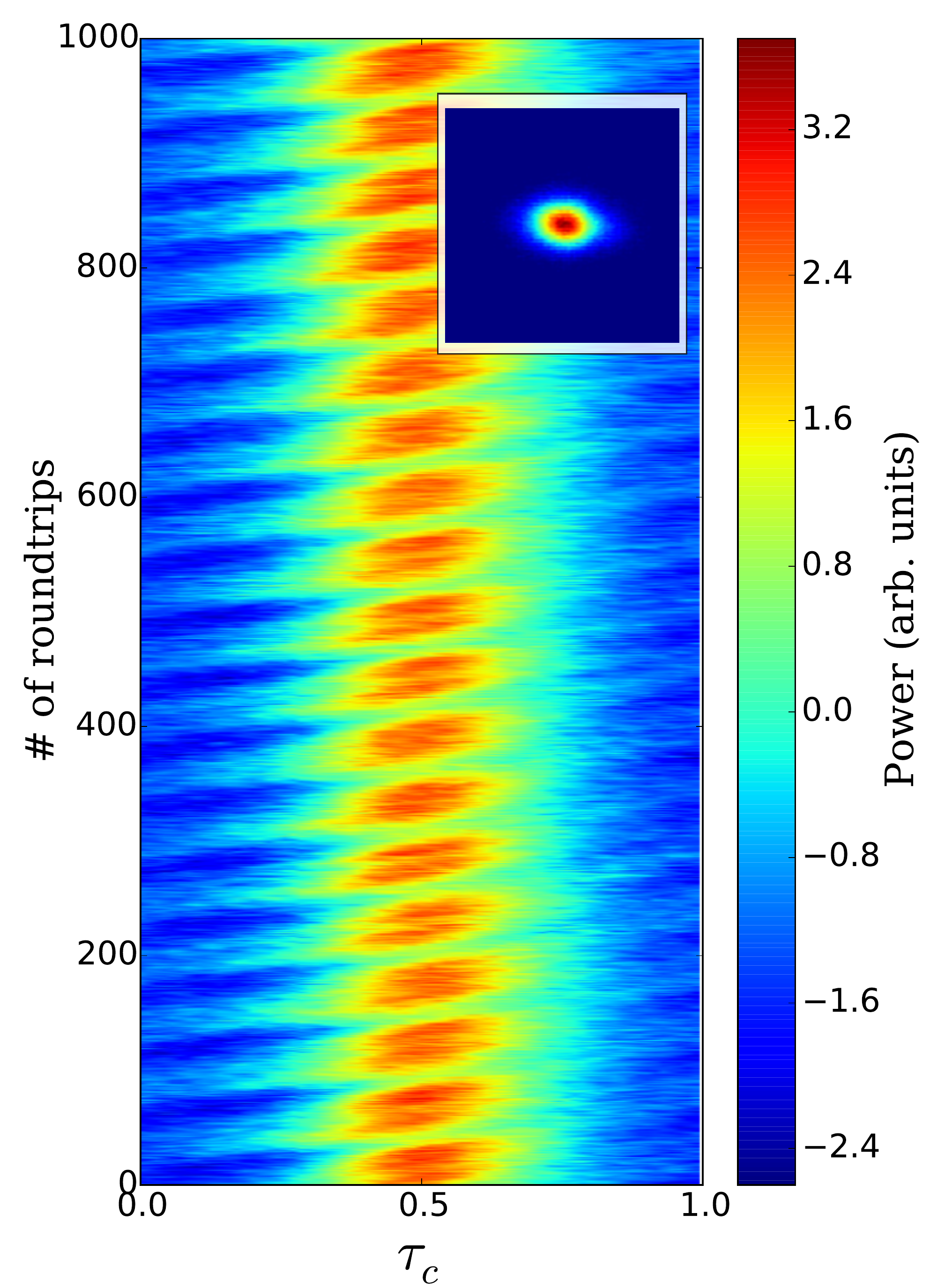}
\protect\caption{(Color online) Example of quasiperiodic dynamics in mode-locking, $I=410.24$~mA.
Explanation of subpanels as in Fig.~\ref{fig:HML}.}
\label{fig:FML_envelope}
\end{figure}

The sidebands are undamped for lower currents (e.g.\ 410.24 mA, Fig.~\ref{fig:FML_envelope})
    %On short time scales there is not much difference to the fundamental mode-locking of Fig.~\ref{fig:FML}.
    %However, on longer time scales the envelope is regularly modulated at low frequencies
leading to a regular modulation of the envelope at low frequencies (27 MHz) as evident from the time series and the space-time
plot. A corresponding peak exists in the low-frequency RF spectrum.
The modulation frequency also appears as a difference frequency between
the sidebands in the RF spectrum around the round-trip frequency where
it is important to note that each of them remains very narrow ($<100$ kHz FWHM).  This corresponds to quasiperiodic motion. Corresponding
instabilities of the envelope of a mode-locked pulse train are known
for Ti:Saphire and fiber lasers \cite{bolton00,soto-crespo04} and can be
described by the cubic-quintic Ginzburg-Landau equation \cite{soto-crespo04,akhmediev01}.

\begin{figure}
\unitlength1mm
\begin{picture}(80,56)
%\centering

\put(0,56){\includegraphics[height=40mm,angle=-90]{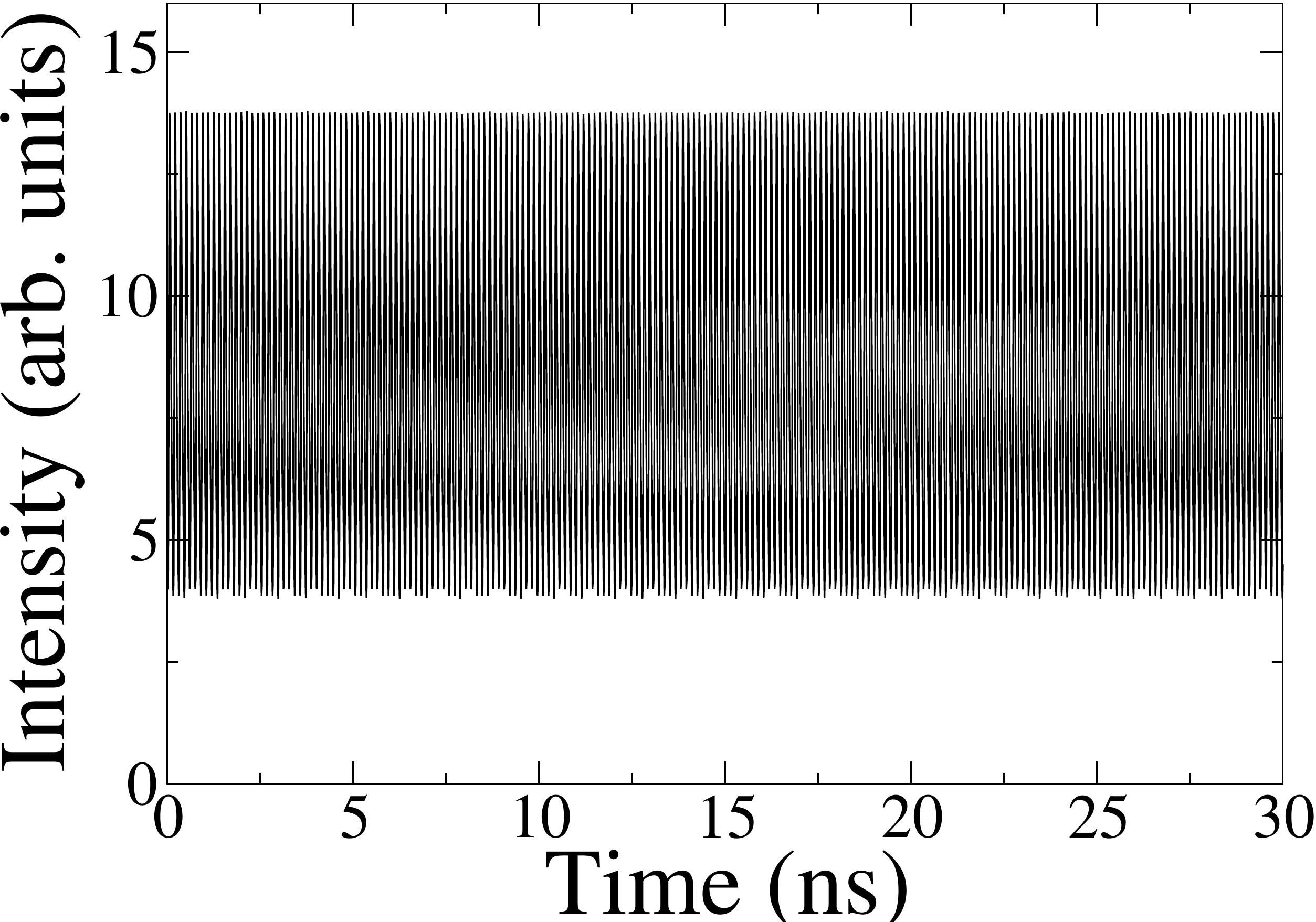}}
\put(0,28){\includegraphics[height=40mm,angle=-90]{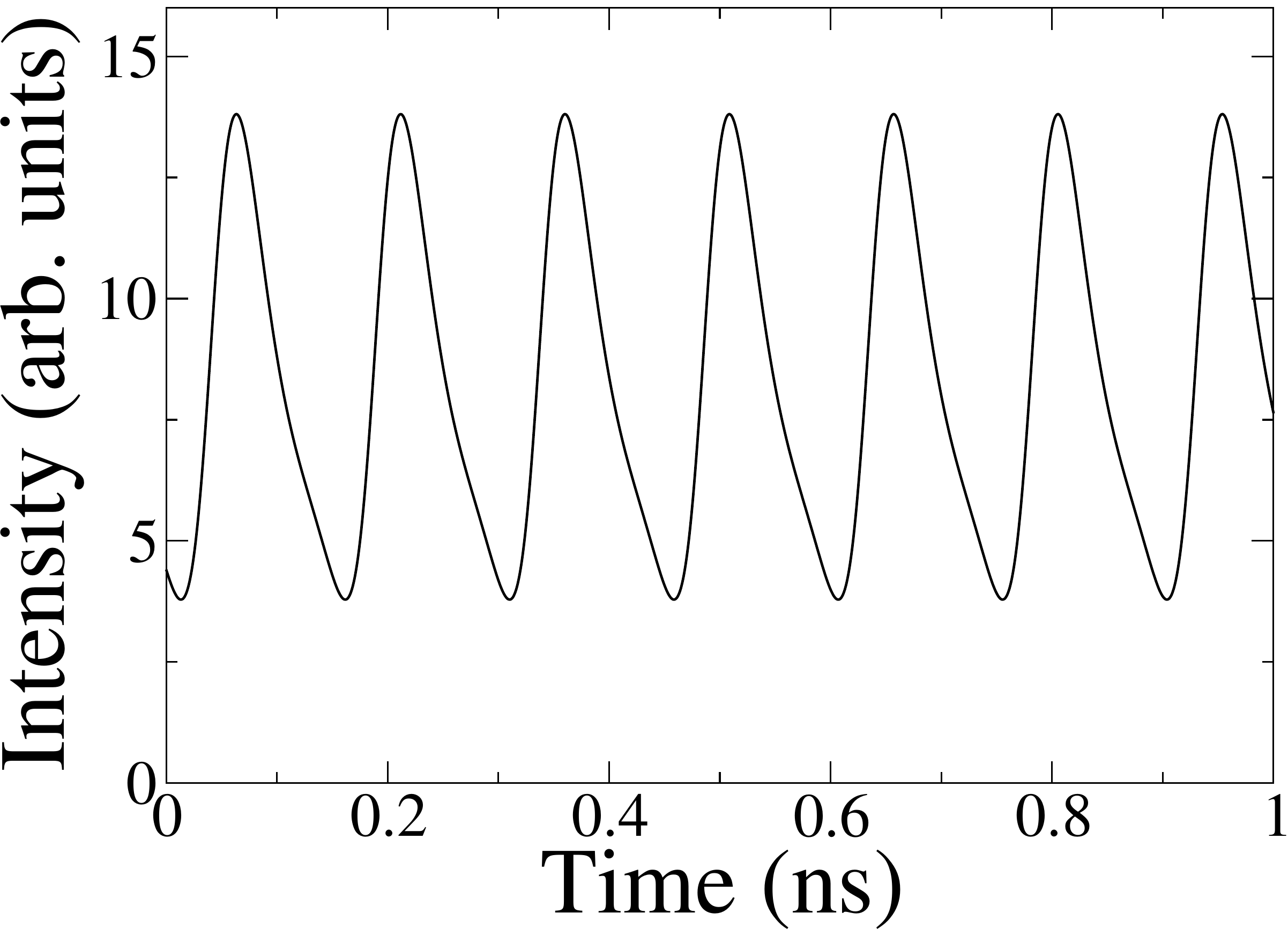}}
\put(40,-1.5){\includegraphics[width=42mm]{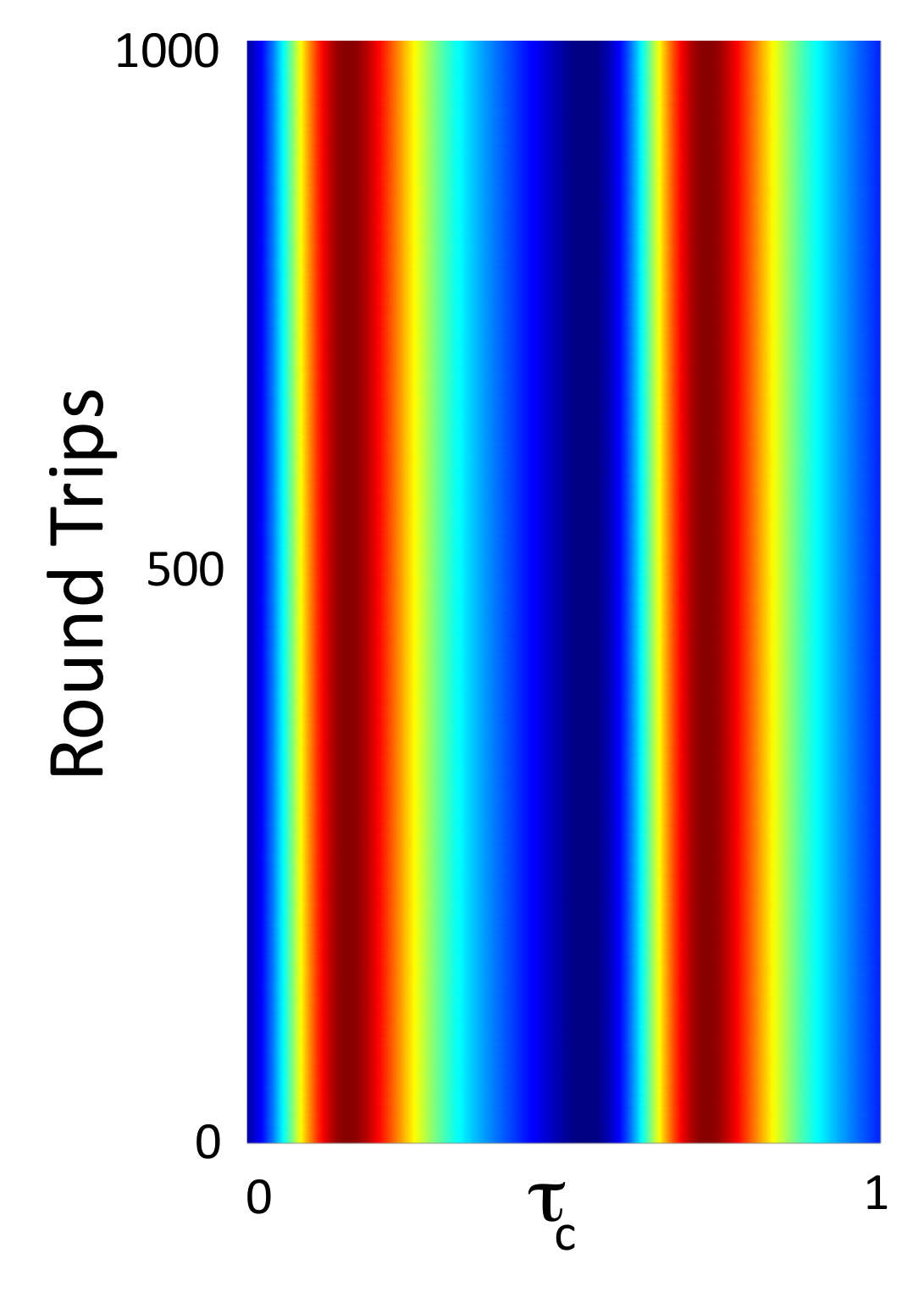}}
\put(66,42){\includegraphics[width=12mm]{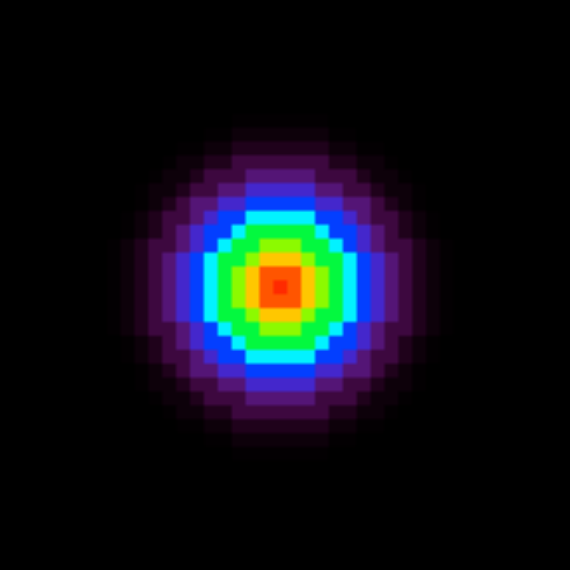}}
\put(8,34){\bf a)}
\put(8,5){\bf b)}
\put(44,50){\bf c)}
\end{picture}
\protect\caption{\label{fig:sim} (Color online)
Numerical results on mode-locked soliton for a cavity round-trip time of 0.28~ns. a) Envelope of time series, b) zoom on pulses, c) space-time representation.
Parameters (see \cite{scroggie09}):  $\alpha=9$, $\theta =-0.95$, $\sigma=0.9$, $\gamma=0.01$, $T_1=0.008$, $T_2=0.002$, $\beta=0.6$, $r_g=0.8$, $\delta=0$, $\tau_f=28$.
The inset shows a snapshot of the intensity profile of the pulsing soliton.
% in a square of approximately 56~$\mu$m size.
% discretization: 0.1953 diffraction length
}
\end{figure}
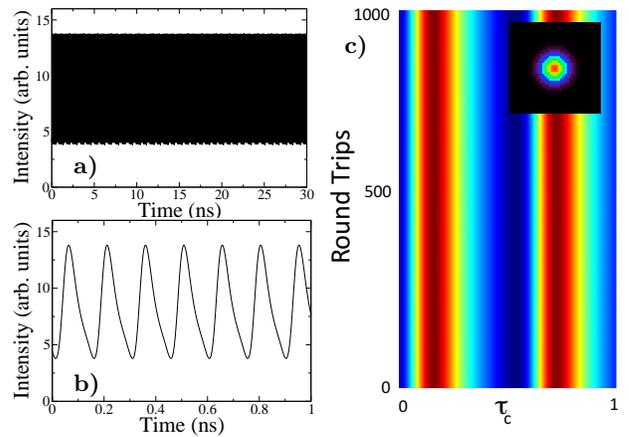

The numerical integration of an established model for the dynamics of a VCSEL with FSF \cite{scroggie09} (see the supplementary material \cite{gustave16s}) confirms the possibility of mode-locked dynamics of spatial solitons. The parameters used are best estimations for the VCSEL under study and were used before to describe cw solitons \cite{scroggie09} and transient pulsing dynamics \cite{radwell10}. Stable mode-locking was found for cavity round-trip times below 0.7 ns with the detuning $\theta$ between the VCSEL and the VBG being a sensitive parameter.  An example of mode-locked dynamics obtained for a cavity round-trip time of 0.28 ns is displayed in Fig.~\ref{fig:sim} showing a pulsing structure very similar to the experimental one. Highly regular pulses are emitted close to the harmonic round-trip frequency. The widths of the first RF-peaks are resolution limited confirming high quality phase correlation. The  FWHM of the pulses is about 40~ps, the modulation depth 78\%. The numerical analysis also confirms that the pulsing is due to an amplitude modulation of the whole profile of the soliton, not a breathing or motional mode.  Cavities with delay times above 0.7~ns yielded irregular dynamics.  In the experiment, stable mode-locking disappeared in longer cavities (1.05~ns round-trip time) in favor of irregular dynamics, too. This constitutes a remarkable qualitative agreement. The large parameter space and the high computational load prevent a more exhaustive analysis of the interplay of simulation parameters like $\alpha-$factor, delay time and carrier lifetime in stable mode-locking at the present stage.

From the observations, we infer  mode-locked dynamics of spatial laser solitons. As the width of the temporal structure is shorter than the cavity round-trip time, this provides evidence for a temporal bullet in the cavity, or,  more carefully, as the modulation is not complete, of a bullet component in the dynamics. A 3D bullet reconstructs itself two times within the round-trip time (four times for harmonic mode-locking), in the VCSEL and at the VBG. This is in line with the fact that,  unless the resonator is completely filled with the nonlinear material, the spatial soliton is only well localized in a few planes within the cavity  \cite{tanguy08,genevet08,taranenko97,saffman94a}. Similarly, large variations of pulse parameters along the cavity might occur for temporal dissipative cavity solitons in lasers which contain discrete gain, loss, and positive and negative dispersion elements \cite{grelu12}.

Self-pulsing instabilities in semiconductor lasers are usually discussed in the framework of the excitation of relaxation oscillations (RO) interacting with the pulsing at the round-trip time \cite{mork90,petermann95}.
For cavities with a round-trip time shorter than the inverse of the RO frequency, fairly regular dynamics in the form of so-called regular pulse packages is possible \cite{heil01,tabaka04,toomey15} but the packages are much shorter than observed here and, more importantly, no stable-self pulsing with a constant envelope was reported in the literature. We do not observe the RO frequency in the output from the VCSEL and the expectation would be that it would exceed the 1.4~GHz external cavity frequency and thus we do not expect the RO to be the dominant driving mechanism.

Looking for the nonlinear mechanisms destabilizing the cw solution, it was realized already some time ago \cite{zorabedian94,trutna93} that a mechanism for nonlinear loss saturation
    %, similar in effect to saturable absorption, nonlinear polarization rotation or Kerr-lensing typically
providing passive mode-locking \cite{grelu12}, might be at work in semiconductor lasers with FSF due to the strong phase-amplitude coupling \cite{henry82}. A reduction of carrier density leads to an increase in refractive index which shifts the resonance of the VCSEL further into resonance with the VBG thus reducing losses. This loss reduction can overcompensate the gain reduction and it was argued to provide a mechanism to amplify fluctuations and multi-longitudinal mode operation. The possibility of regular, slightly anharmonic oscillations with FSF was noticed \cite{zorabedian94}.
    %, but to our knowledge never analyzed in detail.
Four wave mixing mediated by phase-amplitude coupling will also contribute to phase-locking as noted before for semiconductor lasers without saturable absorber section \cite{renaudier05a,gosset06}, soliton formation in dielectric microcavities \cite{herr14} and dissipative parametric instabilities \cite{perego16}. Mechanisms limiting the pulse width are gain dispersion due to the filter bandwidth and the group velocity dispersion introduced by the narrow VCSEL resonance. For the parameters of Fig.~\ref{fig:sim} this can be estimated to be about $-80$~ps$^2$ for the main optical mode taking the effective cavity resonance at the carrier density in the minimum of the pulse as reference to measure detuning. This means that most of the optical spectrum is in the anomalous dispersion regime, i.e.\ in temporal soliton territory, albeit with strong high order dispersion. Note that normal group velocity dispersion would not necessarily exclude the stability of light bullets due to space-time coupling \cite{akhmediev07} and that the existence range of multi-dimensional solitons is typically enhanced in dissipative schemes as argued in the introduction.

Having established a possible mechanism for mode-locking and nonlinear pulse shaping making these structure dissipative solitons in the terminology of \cite{grelu12}, there is the question whether they are truly self-localized states in the sense that they are
    %completely
independent from each other.   This question was recently addressed for temporal solitons in a VCSEL with saturable absorber \cite{marconi14} and a difference between the harmonically mode-locked state, in which solitons are locked to a rigid temporal pattern at maximum distance by the need for gain recovery, and independent localized structures with,  in principle, arbitrary positioning was established. We don't have the equipment for addressing pulses at the sub-cavity round-trip time scale but
    %visible
inspection of Fig.~\ref{fig:HML} indicates that there might be some jitter of pulse to pulse separations. A quantitative analysis of the SDV of the pulse separations yields  a slightly larger jitter between adjacent pulses than between pulses separated by the round-trip time \cite{gustave16s}. One needs to be careful as the tendency is at the edge of statistical significance, but these observations open the tantalizing prospect that the pulses not only represent mode-locked spatial solitons but pave the way to truly mutually independent cavity light bullets.

\begin{acknowledgments}
We are grateful to Roland J\"{a}ger (Ulm Photonics) for supplying the
devices, Jesus Jimenez and Paul
Griffin for help with the experimental setup, Jiazhu Pan for advice on the analysis of the uncertainty of the jitter data, Svetlana
Gurevich for literature on reaction diffusion systems  and LeCroy and Tektronix
for lending us marvellous fast digitizing oscilloscopes. NR and GMcI gratefully acknowledge support from the EPSRC doctoral training account. The Macquarie group gratefully acknowledges support by a Science and Industry Endowment Fund (SIEF) grant. The collaboration between the British and Australian group
was supported by the Royal Society (London), between the British and
the French group by the Global Engagement Fund of the University of
Strathclyde and the CNRS.
\end{acknowledgments}

%\bibliography{Thorstenlit2}

%

\end{document}